\documentclass[a4paper,10pt,
twocolumn,colorlinks,urlcolor=black,linkcolor=black,citecolor=black,
]{article}
\usepackage{amsmath,amssymb}
\usepackage[dvips]{color}
\usepackage{cite}
\usepackage{hangcaption}
\usepackage{amsmath,amssymb}
\usepackage[dvips]{color}
\usepackage{cite}
\usepackage{eepic}
\usepackage{epic}
\usepackage{hangcaption}
\usepackage{amsmath}
\usepackage{amssymb}
\usepackage{amscd}
\usepackage{amsthm}
\usepackage{multicol}
\usepackage[dvipdfmx,%
 bookmarks=true,%
 bookmarksnumbered=true,%
 colorlinks=true,%
 setpagesize=false,%
 pdfkeywords={TeX; dvipdfmx; hyperref; color;}]{hyperref}
\def\cases{\left\{\begin{array}{ll}}
\def\endcases{\end{array}\right.}

\def\bigtimes{\mathop{\mbox{\Large $\times$}}}
\begin{document}
\setcounter{page}{1}
\twocolumn[
\vskip1.5cm
\begin{center}
{\Large \bf 
Heisenberg uncertainty principle and quantum Zeno effects in the linguistic interpretation of quantum mechanics
}
\vskip0.5cm
{\rm
\large
Shiro Ishikawa
}
\\
\vskip0.2cm
\rm
\it
Department of Mathematics, Faculty of Science and Technology,
Keio University,
\\ 
3-14-1, Hiyoshi, Kouhoku-kuYokohama, Japan.
E-mail:
ishikawa@math.keio.ac.jp
\end{center}
\par
\rm
\vskip0.3cm
\par
\noindent
{\bf Abstract}
\normalsize
\vskip0.5cm
\par
\noindent
Recently we proposed measurement theory (i.e., quantum language, or the linguistic interpretation of quantum mechanics), which is characterized as the linguistic turn of the Copenhagen interpretation of quantum mechanics.
This turn from physics to language does not only extend quantum theory to classical theory but also yield the quantum mechanical world view
(i.e., the (quantum) linguistic world view). Thus, we believe that the linguistic interpretation is the most powerful in all interpretations.
Our purpose is to examine the power of measurement theory, that is, to try to formulate Heisenberg uncertainty principle (particulary, the relation between Ishikawa's formulation and so called Ozawa's inequality) and quantum Zeno effects in the linguistic interpretation. As our conclusions, we must say that our trials do not completely succeed. However, we want to believe that this does not imply that we must abandon our linguistic interpretation.

\vskip2.0cm
]

\par

\def\Cal{\cal}
\def\bigstimes{\text{\large $\: \boxtimes \,$}}

\par
\noindent

\vskip0.2cm
\par
\noindent
\par
\noindent
\section{\large
Measurement Theory (= Quantum language)
}
\subsection{\normalsize
Overview
}

\rm
\par
\par
\noindent

In this section,
we shall mention the overview of measurement theory
(or in short, MT).
\par
\par
\rm

\par
\noindent
\par
It is well known ({\it cf.}{{{}}}{\cite{Neum}}) that
quantum mechanics is
formulated 
in an operator algebra $B(H)$
(i.e.,
an operator algebra
composed of all
bounded linear operators
on a Hilbert space $H$
with the norm
$\|F\|_{B(H)}=\sup_{\|u\|_H = 1} \|Fu\|_H$
)
as follows:
\begin{itemize}
\item[(A)]
$
\quad
\underset{\text{\scriptsize (physics)}}{\text{quantum mechanics}}
\qquad
$
\item[]
$
=
{}
{}
\displaystyle{
{
\mathop{\mbox{[quantum measurement]}}_{\text{\scriptsize (probabilistic interpretation) }}
}
}
{}
{}
+
{}
{}
\displaystyle{
\mathop{
\mbox{
[causality]
}
}_{
{
\mbox{
\scriptsize
(
kinetic
equation)
}
}
}
}
{}
$
\end{itemize}
\par
\noindent
Also, the Copenhagen interpretation due to N. Bohr (et al.) is characterized
as
the guide to the usage of quantum mechanics (A).

\par
Measurement theory
({\it cf.} refs.
{{{}}}{\cite{Ishi2}-\cite{Ishi11}})
is,
by an analogy of the (A), constructed
as the mathematical
theory
formulated
in a certain $C^*$-algebra ${\cal A}$
(i.e.,
a norm closed subalgebra in $B(H)$,
{\it cf.} {{{}}}{\cite{Saka}}
)
as follows:
\par
\noindent
\par
\noindent
\vskip0.2cm
\par
\begin{itemize}
\item[(B)]
$
\quad
\underset{\text{\scriptsize (
language)}}{\text{measurement theory}}
\!
$
\item[]
$
=
\displaystyle{
{
\mathop{\mbox{[measurement]}}_{\text{\scriptsize (Axiom 1 in Section 1.2) }}
}
}
+
\displaystyle{
\mathop{
\mbox{
[causality]
}
}_{
{
\mbox{
\scriptsize
(Axiom 2 in Section 1.2)
}
}
}
}
$
\end{itemize}
\par
\par
\noindent
Note that this theory (B) is not physics but a kind of language
based on
{\lq\lq}the mechanical world view{\rq\rq}
since it is a mathematical generalization of 
quantum mechanics (A).
Thus, our linguistic interpretation is characterized
as
the guide to the usage of Axioms 1 and 2 in (B).

When ${\cal A}=B_c(H)$,
the ${C^*}$-algebra composed
of all compact operators on a Hilbert space $H$,
the (B) is called {quantum measurement theory}
(or,
quantum system theory),
which can be regarded as
the linguistic aspect of quantum mechanics.
Also, when ${\cal A}$ is commutative
(that is, 
when ${\cal A}$ is characterized by $C_0(\Omega)$,
the $C^*$-algebra composed of all continuous 
complex-valued functions vanishing at infinity
on a locally compact Hausdorff space $\Omega$
({\it cf.} {{{}}}{\cite{Saka,Yosi}})),
the (B) is called {classical measurement theory}.

Thus, we have the following classification:

\newpage
\begin{itemize}
\item[(C)]
$
\quad
\underset{\text{\scriptsize }}{\text{measurement theory}}
$
\\
$
\;
=
\left\{\begin{array}{ll}
\underset{\text{\scriptsize (when ${\cal A}=C_0(\Omega)$)}}{\text{[C$_s$]:classical system theory}}
\\
\underset{\text{\scriptsize (when ${\cal A}=B_c (H)$)}
}{\text{[Q$_s$]:quantum system theory}}
\end{array}\right.
$
\end{itemize}
\par
\rm
\par
\noindent
\vskip0.5cm
\noindent
\unitlength=0.5mm
\begin{picture}(200,82)(15,0)
\allinethickness{0.5mm}
\put(90,40){{\oval(150,80)}
\put(-68,20){\mbox{[$C_s$]:classical system$\qquad$[$Q_s$]:quantum system}}}
\put(94,0){\line(0,1){80}}
\put(57,30){{{\oval(64,30)}}}
\put(37,36){
\put(-10,-10){classical mechanics} 
\put(10,0){{\mbox{[$C_m$}]}}
}
\put(128,30){\oval(64,30)}
\put(107,36){
\put(-10,-10){quantum mechanics} 
\put(10,0){{\mbox{[$Q_m$}]}}
}

\put(30,-15){\bf
{\bf Figure 1}. 
\rm
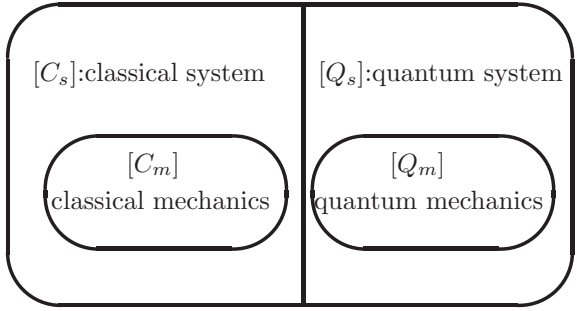
The classification of MT
}
\end{picture}
\vskip1.0cm
\par
\noindent
Thus, this theory covers 
several conventional system theories
(i.e., statistics, dynamical system theory,
quantum system theory).

\par
\noindent
\par
\noindent
\par
\noindent
\subsection{\normalsize
Measurement theory (= Quantum language)
}

\par
\noindent
\par
Measurement theory (B) has two formulations
(i.e.,
the $C^*$-algebraic formulation
and
the $W^*$-algebraic formulation,
{\it cf.} {{{}}}{\cite{Ishi5, Ishi6}}
).
In this paper,
we devote ourselves to
the $W^*$-algebraic formulation
of the measurement theory
(B).

Let
${\cal A}
( \subseteq B(H))$
be
a
${C^*}$-algebra,
and let
${\cal A}^*$ be the
dual Banach space of
${\cal A}$.
$\;\;$
That is,
$ {\cal A }^* $
$ {=}  $
$ \{ \rho \; |$
$ \; \rho$
is a continuous linear functional on ${\cal A}$
$\}$,
and
the norm $\| \rho \|_{ {\cal A }^* } $
is defined by
$ \sup \{ | \rho ({}F{}) |  \:{}\; | \;\; F \in {\cal A}
\text{ such that }\| F \|_{{\cal A}} 
(=\| F \|_{B(H)} )\le 1 \}$.
Define the
\it
mixed state
$\rho \;(\in{\cal A}^*)$
\rm
such that
$\| \rho \|_{{\cal A}^* } =1$
and
$
\rho ({}F) \ge 0
$
for all
$F\in {\cal A}$
such that
$
F \ge 0$.
And define the mixed state space
${\frak S}^m  ({}{\cal A}^*{})$
such that
\begin{align*} {\frak S}^m  ({}{\cal A}^*{})
{=}
\{ \rho \in {\cal A}^*  \; | \;
\rho
\text{ is a mixed state}
\}.
\end{align*}
%
\rm
A mixed state
$\rho (\in {\frak S}^m  ({\cal A}^*) $)
is called a
\it
pure state
\rm
if
it satisfies that
{\lq\lq $\rho = \theta \rho_1 + ({}1 - \theta{}) \rho_2$
for some
$ \rho_1 , \rho_2 \in {\frak S}^m  ({\cal A}^*)$
and
$0 < \theta < 1 $\rq\rq}
implies
{\lq\lq $\rho =  \rho_1 = \rho_2$\rq\rq}\!.
Put
\begin{align*} {\frak S}^p  ({}{\cal A}^*{})
{=}
\{ \rho \in {\frak S}^m  ({\cal A}^*)  \; | \;
\rho
\text{ is a pure state}
\},
\end{align*}
which is called a
\it
state space.
\rm
\rm
It is well known
({\it cf.} {{{}}}{\cite{Saka}})
that
$ {\frak S}^p  ({}{B_c(H)}^*{})=$
$\{ | u \rangle \! \langle u |
$
(i.e., the Dirac notation)
$
\:\;|\;\:
$
$
\|u \|_H=1 
\}$,
and
$ {\frak S}^p  ({}{C_0(\Omega)}^*{})$
$=$
$\{ \delta_{\omega_0} \;|\; \delta_{\omega_0}$ is a point measure at
${\omega_0}
\in \Omega
\}$,
where
$ 
\int_\Omega f(\omega) \delta_{\omega_0} (d \omega )$
$=$
$f({\omega_0})$
$
(\forall f
$
$
\in C_0(\Omega))$.
The latter implies that
$ {\frak S}^p  ({}{C_0(\Omega)}^*{})$
can be also identified with
$\Omega$
(called a {\it spectrum space}
or simply
{\it spectrum})
such as
\begin{align}
\underset{\text{\scriptsize (state space)}}{{\frak S}^p  ({}{C_0(\Omega)}^*{})}
\ni \delta_{\omega} \leftrightarrow {\omega} \in 
\underset{\text{\scriptsize (spectrum)}}{\Omega}
\label{eq1}
\end{align}

\par
\vskip0.2cm
\par
\rm

\par
\par
Consider the pair
$[{\cal A},{\cal N}]_{B(H)}$,
called a
{\it
basic structure}.
Here,
${\cal A} ( \subseteq B(H))$
is a $C^*$-algebra,
and
${\cal N}$
(${\cal A} \subseteq {\cal N} \subseteq B(H)$)
is a particular $C^*$-algebra
(called a $W^*$-algebra)
such that
${\cal N}$ is the weak closure of
${\cal A}$
in $B(H)$.
Let
${\cal N}_*$
be the pre-dual Banach space.

For example,
we see ({\it cf.} {{{}}}{\cite{Saka}}) that,
when ${\cal A}=B_c(H)$,
\begin{itemize}
\item[(i)]
${\cal A}^*=Tr(H)$ (=trace class),
${\cal N}=B(H)$,
${\cal N}_*=Tr(H)$.
\end{itemize}
Also, when ${\cal A}=C_0(\Omega)$,
\begin{itemize}
\item[(ii)]
${\cal A}^*=${\lq\lq}the space of all signed measures on $\Omega$",
${\cal N}=L^\infty ( \Omega, \nu)
(\subseteq B(L^2 ( \Omega, \nu)))$,
${\cal N}_*=L^1 ( \Omega, \nu)$,
where $\nu$ is some measure on $\Omega$
({\it cf.} {{{}}}{\cite{Saka}}).
\end{itemize}

For instance,
in the above (ii) we must clarify the meaning
of the {\lq\lq}value"
of
$F(\omega_0)$
for
$F \in L^\infty(\Omega, \nu )$
and
$\omega_0 \in \Omega$.
An element
$F (\in {\cal N} )$
is
said to be
{\it
essentially continuous
at
}
$\rho_0 (\in  {\frak S}^p  ({}{\cal A}^*{}))$,
if there uniquely exists a complex number $\alpha$
such that
\begin{itemize}
\item[(D)]
if $\rho$
$( \in {\cal N}_*$,
$\| \rho \|_{{\cal N}_*}$
$=1$)
converges to
$\rho_0 (\in  {\frak S}^p  ({}{\cal A}^*{}))$
in the sense of weak$^*$ topology of ${\cal A}^*$,
that is,
\begin{align}
\rho(G) 
\xrightarrow[\quad]{} \rho_0 (G)
\;\;
(\forall G \in {\cal A} (\subseteq {\cal N} )
),
\label{eq2}
\end{align}
then
$\rho(F)$
converges to
$\alpha $.
\end{itemize}
And the value of $\rho_0(F)$ is defined by the $\alpha$.

According to the noted idea (cf. \cite{Davi}),
an {\it observable}
${\mathsf O}{\; :=}(X, {\cal F},$
$F)$ in 
${{\cal N}}$
is defined as follows:
\par
\par
\begin{itemize}
\item[(i)]
[$\sigma$-field]
$X$ is a set,
${\cal F}
(\subseteq 2^X$,
the power set of $X$)
is a $\sigma$-field of $X$,
that is,
{\lq\lq}$\Xi_1, \Xi_2,... \in {\cal F}\Rightarrow \cup_{n=1}^\infty \Xi_n \in {\cal F}$",
{\lq\lq}$\Xi  \in {\cal F}\Rightarrow X \setminus \Xi \in {\cal F}$".
\item[(ii)]
[Countable additivity]
$F$ is a mapping from ${\cal F}$ to ${{\cal N}}$ 
satisfying:
(a):
for every $\Xi \in {\cal F}$, $F(\Xi)$ is a non-negative element in 
${{\cal N}}$
such that $0 \le F(\Xi) $
$\le I$, 
(b):
$F(\emptyset) = 0$ and 
$F(X) = I$,
where
$0$ and $I$ is the $0$-element and the identity
in ${\cal N}$
respectively.
(c):
for any countable decomposition $\{\Xi_1,\Xi_2,\dots ,\Xi_n, ... \}$ 
of 
$\Xi$ $\big($i.e.,
$\Xi,\Xi_n \in {\cal F}\;(n=1,2,3,...)$, 
$\cup_{n=1}^\infty \Xi_n = \Xi$, 
$\Xi_i \cap \Xi_j = \emptyset$ $(i \ne j) \big)$, it holds that
$
F(\Xi)
$
$
=
$
$
\sum_{n=1}^\infty  F(\Xi_n)
$
in the sense of weak$^*$ topology in ${\cal N}$.
\end{itemize}

\par
\vskip0.2cm
\par
\rm
With any {\it system} $S$, a 
basic structure
$[{\cal A},{\cal N}]_{B(H)}$
%
can be associated in which the 
measurement theory (B) of that system can be formulated.
A {\it state} of the system $S$ is represented by an element
$\rho (\in {\frak S}^p  ({}{\cal A}^*{}))$
and an {\it observable} is represented by an observable 
${\mathsf{O}}{\; :=} (X, {\cal F}, F)$ in ${{\cal N}}$.
Also, the {\it measurement of the observable ${\mathsf{O}}$ for the system 
$S$ with the state $\rho$}
is denoted by 
${\mathsf{M}}_{\cal N} ({\mathsf{O}}, S_{[\rho]})$
$\big($
or more precisely,
${\mathsf{M}}_{\cal N} ({\mathsf{O}}{\; :=} (X, {\cal F}, F),$
$ S_{[\rho]})$
$\big)$.
An observer can obtain a measured value $x $
($\in X$) by the measurement 
${\mathsf{M}}_{\cal N} ({\mathsf{O}}, S_{[\rho]})$.
\par
\noindent
\par
The Axiom 1 presented below is 
a kind of mathematical generalization of Born's probabilistic interpretation of quantum mechanics (A).
And thus, it is a statement without reality.

\par
Now we can present Axiom 1 in the $W^*$-algebraic formulation as follows.
\par
\noindent
\bf
Axiom 1
\rm
[
Measurement
].
\it
The probability that a measured value $x$
$( \in X)$ obtained by the measurement 
${\mathsf{M}}_{\cal N} ({\mathsf{O}}{\; :=} (X, {\cal F}, F),$
$ S_{[\rho_0]})$
%
belongs to a set 
$\Xi (\in {\cal F})$ is given by
$
\rho_0( F(\Xi) )
$
if
$F(\Xi)$
is essentially continuous at $\rho_0 ( \in {\frak S}^p  ({}{\cal A}^*{}) )$.
\rm
\par
\vskip0.3cm

\par
Next, we explain Axiom 2.
Let
$[{\cal A}_1,{\cal N}_1]_{B(H_1)}$
and
$[{\cal A}_2,{\cal N}_2]_{B(H_2)}$
be basic structures.
A continuous
linear operator
$\Phi_{1,2}$
$:{{{{\cal N}_2}}}$
(with weak$^*$ topology)
$\to {{{{\cal N}_1}}}$(with weak$^*$ topology)
is called a
\it
Markov operator,
\rm
if
it satisfies that
(i):
$\Phi_{1,2} (F_2) \ge 0$
for any non-negative element
$F_2$ in
${{{{\cal N}_2}}}$,
(ii):
$ \Phi_{1,2}({}I_2{}) = I_1 $,
where $I_k$
is the identity in ${\cal N}_k$,
$(k=1,2)$.
In addition to the above (i) and (ii),
in this paper
we
assume that
$\Phi_{1,2}({\cal A}_2) \subseteq {\cal A}_1$
and
$\sup \{ \|\Phi_{1,2}( F_2) \|_{{\cal A}_1}
\; |
\;
F_2 \in {\cal A}_2
\text{ such that }
\|F_2\|_{{\cal A}_2} \le 1
\}
=1$.

It is clear that
the dual operator
$ \Phi_{1,2}^*{}: $
$
{\cal A}_{1}^*
\to {\cal A}_{2}^*
$
satisfies that
$ \Phi_{1,2}^*{}
(
{\frak S}^m  ({\cal A}_{1}^*)
)
\subseteq
{\frak S}^m  ({\cal A}_{2}^*)
$.

Here note that,
for any observable
${\mathsf{O}}_2{\; :=}({}X , {\cal F}, {F}_2{})$
in ${{{{\cal N}_2}}}$,
the $({}X , {\cal F}, $
$\Phi_{1,2} F_2 )$
is an observable in ${{{\cal N}_1}}$.

%


\par
\vskip0.3cm

\par
\par
Let $(T,\le)$ be a tree, i.e., a partial ordered 
set such that {\lq\lq$t_1 \le t_3$ and $t_2 \le t_3$\rq\rq} implies {\lq\lq$t_1 \le t_2$ or $t_2 \le t_1$\rq\rq}\!.
Put $T^2_\le = \{ (t_1,t_2) \in T^2{}\;|\; t_1 \le t_2 \}$.
Here, note that $T$ is not necessarily finite.

Assume the completeness of the ordered set $T$.
That is,
for any subset $T'( \subseteq T)$
bounded from below
(i.e.,
there exists $t' (\in T)$
such that $t' \le t$
$( \forall t \in T' )$),
there uniquely exists
an element
$\text{inf} (T')$
$\in T$
satisfying the following conditions,
(i):
${\text{inf}} ( T' ) {{\; \leqq \;}}t \; ( \forall t \in T' )$,
(ii):
if
$s {{\; \leqq \;}}t \; \; ( \forall t \in T' )$,
then
$s {{\; \leqq \;}}{\text{inf}} ( T' )$.

\par
\noindent
\par
\noindent
\par
The family
$\{ \Phi_{t_1,t_2}{}: $
${\cal N}_{t_2} \to {\cal N}_{t_1} \}_{(t_1,t_2) \in T^2_\le}$
is called a {\it Markov relation}
({\it due to the Heisenberg picture}),
\rm
if it satisfies the following conditions {\rm (i) and (ii)}.
\begin{itemize}
\item[{\rm (i)}]
With each
$t \in T$,
a basic structure
$[{\cal A}_t,{\cal N}_t]_{B(H_t)}$
is associated.
\item[{\rm (ii)}]
For every $(t_1,t_2) \in T_{\le}^2$, a Markov operator 
$\Phi_{t_1,t_2}{}: {\cal N}_{t_2} \to {\cal N}_{t_1}$ 
is defined.
And it satisfies that
$\Phi_{t_1,t_2} \Phi_{t_2,t_3} = \Phi_{t_1,t_3}$ 
holds for any $(t_1,t_2)$, $(t_2,t_3)$
$ \in$
$ T_\le^2$.
\end{itemize}
\noindent

%
%
When
$ \Phi_{t_1,t_2}^*{}$
$
(
{\frak S}^p  ({\cal A}_{t_1}^*)
)$
$\subseteq
$
$
(
{\frak S}^p  ({\cal A}_{t_2}^*)
)$
holds for any
$
{(t_1,t_2) \in T^2_\le}$,
the Markov relation is said to be
deterministic.
Note that
the classical deterministic Markov relation
is represented by
$\{ \phi_{t_1,t_2}{}: $
${\Omega}_{t_1} \to {\Omega}_{t_2} \}_{(t_1,t_2) \in T^2_\le}$,
where
the continuous map
$\phi_{t_1,t_2}{}: $
${\Omega}_{t_1} \to {\Omega}_{t_2}$
is defined by
\begin{align*}
\Phi_{t_1,t_2}^* (\delta_{\omega_1} )
=
\delta_{
\phi_{t_1,t_2} (\omega_1)}
\quad
(\forall \omega_1 \in \Omega_1)
\end{align*}

\par
\par
\rm
Now Axiom 2 is presented
as follows:
\rm
\par
\noindent
\bf
Axiom 2
\rm
[Causality].
\it
The causality is represented by
a Markov relation 
$\{ \Phi_{t_1,t_2}{}: $
${\cal N}_{t_2} \to {\cal N}_{t_1} \}_{(t_1,t_2) \in T^2_\le}$.

\rm

\par

\par
\noindent
\subsection{\normalsize
The Linguistic Interpretation
}
Next,
we have to
study how to use the above axioms
as follows.
That is, we present the following interpretation
(E)
[=(E$_1$)--(E$_3$)],
which is characterized as a kind of linguistic turn
of so-called Copenhagen interpretation
({\rm cf.}\textcolor{black}{\cite{Ishi6, Ishi7, Ishi8}}
).
That is,
we propose:
\begin{itemize}
\item[(E$_1$)]
Consider the dualism composed of {\lq\lq}observer{\rq\rq} and {\lq\lq}system( =measuring object){\rq\rq}.
And therefore,
{\lq\lq}observer{\rq\rq} and {\lq\lq}system{\rq\rq}
must be absolutely separated. In this sense, the interaction (or, measurement process such as $\textcircled{a}$ and $\textcircled{b}$ In Figure 2) should not be mentioned explicitly.
\end{itemize}

\par
\noindent
\vskip0.2cm
\par
\rm
\par
\noindent
\vskip0.5cm
\noindent
\unitlength=0.5mm
\begin{picture}(200,82)(15,0)
\put(-8,0)
{
\allinethickness{0.2mm}
\drawline[-40](80,0)(80,62)(30,62)(30,0)
\drawline[-40](130,0)(130,62)(175,62)(175,0)
\allinethickness{0.5mm}
\path(20,0)(175,0)
%
\put(14,-5){
\put(37,50){$\bullet$}
}
\put(50,25){\ellipse{17}{25}}
\put(50,44){\ellipse{10}{13}}
\put(0,44){\put(43,30){\sf \footnotesize{observer}}
\put(42,25){\scriptsize{(I(=mind))}}
}
\put(7,7){\path(46,27)(55,20)(58,20)}
\path(48,13)(47,0)(49,0)(50,13)
\path(51,13)(52,0)(54,0)(53,13)
\put(0,26){
\put(142,48){\sf \footnotesize system}
\put(143,43){\scriptsize (matter)}
}
\path(152,0)(152,20)(165,20)(150,50)(135,20)(148,20)(148,0)
\put(10,0){}
\allinethickness{0.2mm}
\put(0,-5){
\put(130,39){\vector(-1,0){60}}
\put(70,43){\vector(1,0){60}}
\put(92,56){\sf \scriptsize \fbox{observable}}
\put(58,50){\sf \scriptsize }
\put(57,53){\sf \scriptsize \fbox{\shortstack[l]{measured \\ value}}}
\put(80,44){\scriptsize \textcircled{\scriptsize a}interfere}
\put(80,33){\scriptsize \textcircled{\scriptsize b}perceive a reaction}
\put(130,56){\sf \scriptsize \fbox{state}}
}
}
\put(30,-15){\bf
{\bf Figure 2}. 
\rm
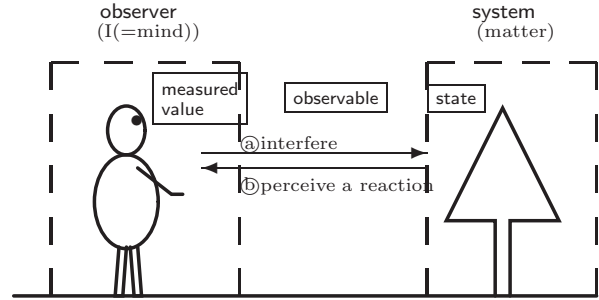
Dualism
in MT
(cf. \cite{Ishi6, Ishi7})
}
\end{picture}
\vskip1.0cm
\par
\noindent

\begin{itemize}
\item[(E$_2$)]
Only one measurement is permitted.
And thus,
the state after a measurement
is meaningless
$\;$
since it 
can not be measured any longer.
Thus, the collapse of the wavefunction is prohibited.  
We are not concerned with anything after measurement. Also, the causality should be assumed only in the side of system,
however,
a state never moves.
Thus,
the Heisenberg picture should be adopted,
and thus,
the Schr\"{o}dinger picture
should be prohibited.
\item[(E$_3$)]
Also, the observer
does not have
the space-time.
Thus, 
the question:
{\lq\lq}When and where is a measured value obtained?{\rq\rq}
is out of measurement theory.
And thus,
Schr\"{o}dinger's cat is out of measurement theory,
\end{itemize}
\par
\noindent
and so on.

And therefore, in spite of
Bohr's realistic view,
we propose the following linguistic view:
\begin{itemize}
\item[(F$_1$)]
In the beginning was the language called measurement theory
(with the linguistic interpretation (E)).
And, for example, quantum mechanics can be
fortunately
described in this language.
And moreover,
almost all scientists have already mastered this language
partially and informally
since statistics
(at least, its basic part)
is characterized as one of aspects of
measurement theory
({\rm cf.} \cite{Ishi4,Ishi9,Ishi10, Ishi11}).
\end{itemize}
For completeness, we again note that,
\begin{itemize}
\item[(F$_2$)]
The linguistic interpretation (E$_1$)-(E$_3$) is not only applied to quantum mechanics ([Q$_m$] in Figure 1) but also to whole MT.
\end{itemize}
This generalization has a merit such as the linguistic interpretation is determined 
"uniquely", though so called Copenhagen interpretation has various  variations.
For example, the projection postulate of quantum mechanics can not be naturally extended to MT.
%
%
%
%

\par
\noindent
\subsection{\normalsize
Sequential Causal Observable
and Its Realization
}

\par
For each
$k=1,$
$2,\ldots,K$,
consider a measurement
${\mathsf{M}}_{{{\cal N}}} ({\mathsf{O}_k}$
${\; \equiv} (X_k, {\cal F}_k, F_k),$
$ S_{[\rho]})$.
However,
since
the (E$_2$)
says that
only one measurement is permitted,
the
measurements
$\{
{\mathsf{M}}_{{{\cal N}}} ({\mathsf{O}_k},S_{[\rho]})
\}_{k=1}^K$
should be reconsidered in what follows.
Under the commutativity condition such that
\begin{align}
&
F_i(\Xi_i) F_j(\Xi_j) 
=
F_j(\Xi_j) F_i(\Xi_i)
\label{eq3}
%
\\
&
\quad
(\forall \Xi_i \in {\cal F}_i,
\forall \Xi_j \in  {\cal F}_j , i \not= j),
\nonumber
\end{align}
we can
define the product observable
${\text{\large $\times$}}_{k=1}^K {\mathsf{O}_k}$
$=({\text{\large $\times$}}_{k=1}^K X_k ,$
$ \boxtimes_{k=1}^K {\cal F}_k,$
$ 
{\text{\large $\times$}}_{k=1}^K {F}_k)$
in ${\cal N}$ such that
\begin{align*}
({\text{\large $\times$}}_{k=1}^K {F}_k)({\text{\large $\times$}}_{k=1}^K {\Xi}_k )
=
F_1(\Xi_1) F_2(\Xi_2) \cdots F_K(\Xi_K)
\\
\;
(
\forall \Xi_k \in {\cal F}_k,
\forall k=1,\ldots,K
).
\qquad
\qquad
\nonumber
\end{align*}
Here,
$ \boxtimes_{k=1}^K {\cal F}_k$
is the smallest field including
the family
$\{
{\text{\large $\times$}}_{k=1}^K \Xi_k
$
$:$
$\Xi_k \in {\cal F}_k \; k=1,2,\ldots, K \}$.
Then, 
the above
$\{
{\mathsf{M}}_{{{\cal N}}} ({\mathsf{O}_k},S_{[\rho]})
\}_{k=1}^K$
is,
under the commutativity condition \textcolor{black}{(3)},
represented by the simultaneous measurement
${\mathsf{M}}_{{{{\cal N}}}} (
{\text{\large $\times$}}_{k=1}^K {\mathsf{O}_k}$,
$ S_{[\rho]})$.

\par
Consider a tree
$(T{\; \equiv}\{t_0, t_1, \ldots, t_n \},$
$ \le )$
with the root $t_0$.
This is also characterized by
the map
$\pi: T \setminus \{t_0\} \to T$
such that
$\pi( t)= \max \{ s \in T \;|\; s < t \}$.
Let
$\{ \Phi_{t, t'} : {\cal N}_{t'}  \to {\cal N}_{t}  \}_{ (t,t')\in
T_\le^2}$
be a causal relation,
which is also represented by
$\{ \Phi_{\pi(t), t} : {\cal N}_{t}  \to {\cal N}_{\pi(t)}  \}_{ 
t \in T \setminus \{t_0\}}$.
Let an observable
${\mathsf O}_t{\; \equiv}
(X_t, {\cal F}_{t}, F_t)$ in the ${\cal N}_t$ 
be given for each $t \in T$.
Note that
$\Phi_{\pi(t), t}
{\mathsf O}_t$
$(
{\; \equiv}
(X_t, {\cal F}_{t},
\Phi_{\pi(t), t} F_t)$
)
is an observable in the ${\cal N}_{\pi(t)}$.

The pair
$[{\mathbb O}_T]
$
$=$
$[
\{{\mathsf O}_t \}_{t \in T}$,
$\{ \Phi_{t, t'} : {\cal N}_{t'}  \to {\cal N}_{t}  \}_{ (t,t')\in
T_\le^2}$
$]$
is called a
{\it sequential causal observable}.
For each $s \in T$,
put $T_s =\{ t \in T \;|\; t \ge s\}$.
And define the observable
${\widehat{\mathsf O}}_s
\equiv ({\text{\large $\times$}}_{t \in T_s}X_t, \boxtimes_{t \in T_s}{\cal F}_t, {\widehat{F}}_s)$
in ${\cal N}_s$
as follows:
\par
\noindent
\begin{align}
\widehat{\mathsf O}_s
&=
\left\{\begin{array}{ll}
{\mathsf O}_s
\quad
&
\!\!\!\!\!\!\!\!\!\!\!\!\!\!\!\!\!\!
\text{(if $s \in T \setminus \pi (T) \;${})}
\\
{\mathsf O}_s
{\text{\large $\times$}}
({}\bigtimes_{t \in \pi^{-1} ({}\{ s \}{})} \Phi_{ \pi(t), t} \widehat {\mathsf O}_t{})
\quad
&
\!\!\!\!\!\!
\text{(if $ s \in \pi (T) ${})}
\end{array}\right.
\label{eq4}
\end{align}
if
the commutativity condition holds
(i.e.,
if the product observable
${\mathsf O}_s
{\text{\large $\times$}}
({}\bigtimes_{t \in \pi^{-1} ({}\{ s \}{})} \Phi_{ \pi(t), t}
$
$\widehat {\mathsf O}_t{})$
exists)
for each $s \in \pi(T)$.
Using \textcolor{black}{(4)} iteratively,
we can finally obtain the observable
$\widehat{\mathsf O}_{t_0}$
in ${\cal N}_{t_0}$.
The
$\widehat{\mathsf O}_{t_0}$
is called the realization
(or,
realized causal observable)
of
$[{\mathbb O}_T]$.

\par
\par
\noindent
\subsection{\normalsize
Our motivation
}
Since MT has been shown to have a great power (cf. \cite{Ishi2}-\cite{Ishi11}), we believe that MT is superior to statistics as the language of science.
However we now feel misgivings about the possibility that Heisenberg uncertainty principle (particularly, Ozawa's inequality(cf. \cite{Oza, Oza2}) and quantum Zeno effects (cf. \cite{Misr}) can not be formulated in MT.  As our conclusions, we say that our trials partially succeed. This may imply that these should be understood in the other interpretations.

\par
\noindent
\section{\large
Heisenberg uncertainty principle
}
\par
\noindent
\subsection{\normalsize
Preliminary
}
Let
$[B_c(H),B(H)]_{B(H)}$ 
be the basic structure.
Let $A_i$ $(i=1,2)$ be arbitrary self-adjoint operator on $H$. For example, it may satisfy that $[A_1 , A_2](:=A_1 A_2 - A_2 A_1 ) =\hbar \sqrt{-1}I$.
Let ${\mathbb R}$ and ${\cal B}$ be the real line and its Borel field respectively.
Let ${\mathsf O}_{A_i}=({\mathbb R}, {\cal B}, F_{A_i} )$ be the spectral representation of $A_i$, i.e., $A_i=\int_{\mathbb R} \lambda F_{A_i}( d \lambda )$, which is regarded as the observable in $B(H)$.
Let $\rho_u= |u\rangle  u |$ be a state, where $u \in H$ and $\|u\|=1$.
Thus, we two measurements:
\par
\noindent
\begin{itemize}
\item[(G$_1$)]${\mathsf{M}}_{B(H)} ({\mathsf{O}_{A_1}}{\; :=} ({\mathbb R}, {\cal B}, F_{A_1} ),$
$ S_{[\rho_u]})$
\item[(G$_2$)]${\mathsf{M}}_{B(H)} ({\mathsf{O}_{A_2}}{\; :=} ({\mathbb R}, {\cal B}, F_{A_2} ),$
$ S_{[\rho_u]})$
\end{itemize}

Let $K$ be another Hilbert space, and let $s$ be in $K$ such that $\| s \|=1$. Thus, we also have two observables ${\mathsf{O}_{A_1 \otimes I}}{\; :=} ({\mathbb R}, {\cal B}, F_{A_1} \otimes I )$ and ${\mathsf{O}_{A_2\otimes I}}{\; :=} ({\mathbb R}, {\cal B}, F_{A_2}\otimes I )$ in $B(H \otimes K)$, where $I \in B(K)$ is the identity map.  Put ${\widehat \rho}_{us}=|u \otimes s \rangle \langle u \otimes s|$.
Here, we have two measurements as follows:
\par
\noindent
\begin{itemize}
\item[(H$_1$)]${\mathsf{M}}_{B(H\otimes K)} ({\mathsf{O}_{A_1 \otimes I}},S_{[{\widehat \rho}_{us}]})$
\item[(H$_2$)]${\mathsf{M}}_{B(H\otimes K)} ({\mathsf{O}_{A_2 \otimes I}},S_{[{\widehat \rho}_{us}]})$
\end{itemize}
which is clearly equivalent to the above two (G$_1$) and (G$_2$) respectively

Now we want to take these two measurements. However, the linguistic interpretation (E$_2$) says that it is impossible, if $A_1$ and $A_2$ do not commute.

Let ${\widehat A}_i$ $(i=1,2)$ be arbitrary self-adjoint operator on the tensor Hilbert space $H \otimes K$, where it is assumed that
\begin{align}
[{\widehat A}_1, {\widehat A}_2](:=
{\widehat A}_1{\widehat A}_2- {\widehat A}_2{\widehat A}_1)=0
\label{eq5}
\end{align}
Let ${\mathsf O}_{{\widehat A}_i}=({\mathbb R}, {\cal B},
F_{{\widehat A}_i} )$ be the spectral representation of ${\widehat A}_i$, i.e.${\widehat A}_i=\int_{\mathbb R} \lambda F_{{\widehat A}_i}
( d \lambda )$, which is regarded as the observable in $B(H \otimes K)$.
Thus, we have two measurements as follows:
\par
\noindent
\begin{itemize}
\item[(I$_1$)]${\mathsf{M}}_{B(H\otimes K)} ({\mathsf{O}_{{\widehat A}_1}},S_{[{\widehat \rho}_{us}]})$
\item[(I$_2$)]${\mathsf{M}}_{B(H\otimes K)} ({\mathsf{O}_{{\widehat A}_2}},S_{[{\widehat \rho}_{us}]})$
\end{itemize}
Note, by the commutative condition (\ref{eq5}), that
\it
the two can be realized as the simultaneous measurement
\rm
${\mathsf{M}}_{B(H\otimes K)} ({\mathsf{O}_{{\widehat A}_1}}\times{\mathsf{O}_{{\widehat A}_2}},S_{[{\widehat \rho}_{us}]})$, where ${\mathsf{O}_{{\widehat A}_1}}\times{\mathsf{O}_{{\widehat A}_2}}=({\mathbb R}^2, {\cal B}^2,
F_{{\widehat A}_1} \times F_{{\widehat A}_2} )$.

Again note that any relation between $A_i \otimes I$ and ${\widehat A}_i$ is not assumed.
However, 
\it
we want to regard this simultaneous measurement as the substitute of the above two (H$_1$) and (H$_2$).
\rm
Putting 
\begin{align}
{\widehat N}_i := & {\widehat A}_i -A_i \otimes I
\nonumber
\\
(\text{and thus, }&  {\widehat A}_i={\widehat N}_i +A_i \otimes I)
\label{eq6}
\end{align}
we define the $\Delta_{\widehat{N}_i}^{u \otimes s}$ and ${\overline \Delta}_{\widehat{N}_i}^{u \otimes s}$ such that
\begin{align}
&
\Delta_{\widehat{N}_i}^{u \otimes s} =\|  {\widehat N}_i (u \otimes s) \|
\label{eq7}
\\
&
{\overline \Delta}_{\widehat{N}_i}^{u \otimes s} =\| ( {\widehat N}_i - \langle  u \otimes s , {\widehat N}_i (u \otimes s)\rangle ) (u \otimes s) \|
\nonumber 
\end{align}
where the following inequality:
\begin{align}
\Delta_{\widehat{N}_i}^{u \otimes s} 
\ge
{\overline \Delta}_{\widehat{N}_i}^{u \otimes s} 
\label{eq8}
\end{align}
is common sense.

By the commutative condition (\ref{eq5}) and (\ref{eq6}), we see that
\begin{align}
&[{\widehat N}_1,{\widehat N}_2]
+
[{\widehat N}_1, A_2 \otimes I]+[A_1 \otimes I ,{\widehat N}_2]
\nonumber
\\
&
=
-[A_1 \otimes I, A_2 \otimes I]
\label{eq9}
\end{align}

Here, we should note that the first term (or, precisely,
$\langle u \otimes s,$"the first term"$(u \otimes s) \rangle$) of (\ref{eq9}) can be, by the Robertson uncertainty relation ({\it cf.}{{{}}}{\cite{Neum}}), estimated as follows:
\begin{align}
& 
2 {\overline \Delta}_{\widehat{N}_1}^{u \otimes s} \cdot {\overline \Delta}_{\widehat{N}_2}^{u \otimes s}
\nonumber
\\ \ge
&
| \langle u \otimes s ,
[{\widehat N}_1,{\widehat N}_2] ( u \otimes s) \rangle |
\label{eq10}
\end{align}
\par
\noindent
{\bf Remark 1 }There may be an opinion such that the physical meaning of ${\Delta}_{\widehat{N}_1}^{u \otimes s}$ (or, ${\overline \Delta}_{\widehat{N}_1}^{u \otimes s}$) is not clear. However, we do not worry about this problem. That is because our concern is not only quantum mechanics ([Q$_m$] in Figure 1) but also
quantum system theory ([Q$_s$] in Figure 1). However, recalling (F$_2$), in most cases, we can expect that
\begin{itemize}
\item[(J)] A ( metaphysical) statement in quantum system theory is regarded as a (physical) statement in quantum mechanics,
\end{itemize}
because both are formulated in the same mathematical structure, and moreover, are based on the linguistic interpretation.
\par
\noindent
\subsection{\normalsize
Heisenberg uncertainty principle with the same average condition
}

In the previous section, any relation between $A_i \otimes I$ and ${\widehat A}_i$ is not assumed. However, in this section we assume the following hypothesis:
\par
\noindent
{\bf Hypothesis 1} (The same average condition). We assume that
\begin{align}
&
\langle u \otimes s,  {\widehat N}_i(u \otimes s) \rangle =0 \qquad ( \forall u \in H, i=1,2)
\label{eq11}
\end{align}
or equivalently
\begin{align*}
\langle u \otimes s,  {\widehat A}_i(u \otimes s) \rangle =
\langle u ,  {A}_i u \rangle  \quad ( \forall u \in H, i=1,2)
\end{align*}
holds. Thus, in this case, it holds that
\begin{align}
\Delta_{\widehat{N}_i}^{u \otimes s}=
{\overline \Delta}_{\widehat{N}_i}^{u \otimes s}
\label{eq12}
\end{align}

\par
\noindent
{\bf Remark 2} The existence of ${\widehat A}_i$ (with the conditions (\ref{eq5}) and (\ref{eq11})) is guaranteed ({\rm cf.} \cite{Ishi1}).
Also, we can assume that the (\ref{eq11}) is equivalent to
\begin{align}
&
\langle u \otimes s,  {\widehat N}_i(v \otimes s) \rangle =0
\quad ( \forall u, v \in H, i=1,2)
\label{eq13}
\end{align}
This is proved as follows: 
\begin{align*}
0 &=
\langle ( u + v) \otimes s,  {\widehat N}_i((u +v) \otimes s) \rangle 
\\
&
=
\langle  u  \otimes s,  {\widehat N}_i (v \otimes s) \rangle 
+
\langle v \otimes s,  {\widehat N}_i (u  \otimes s) \rangle 
\\
&
=
2 \mbox{[Real part]}(\langle  u  \otimes s,  {\widehat N}_i (v \otimes s) \rangle )
\\
0
&
=\langle ( u + \sqrt{-1} v) \otimes s,  {\widehat N}_i((u +\sqrt{-1} v) \otimes s) \rangle 
\\
&
=
2 \sqrt{-1}  \mbox{[Imaginary part]}(\langle  u  \otimes s,  {\widehat N}_i (v \otimes s) \rangle )
\end{align*}	
Thus we get (\ref{eq13}).
\vspace{6mm}

Using (\ref{eq13}), we can calculate the second term (or, precisely,
$\langle u \otimes s,$"the second term"$(u \otimes s) \rangle$) 
in (\ref{eq9}) as follows:
\begin{align}
&
\langle u \otimes s, [{\widehat N}_1, A_2 \otimes I](u \otimes s) \rangle 
\nonumber
\\
=
& 
\langle u \otimes s, {\widehat N}_1 (A_2 u \otimes s) \rangle 
-
\langle A_2 u \otimes s,  {\widehat N}_1( u \otimes s) \rangle 
\nonumber
\\
=
& 0
\qquad ( \forall u \in H)
\label{eq14}
\end{align}
Similarly, we calculate the third term in (\ref{eq9}) as follows:
\begin{align}
&
\langle u \otimes s, [A_1 \otimes I, {\widehat N}_2](u \otimes s) \rangle =0
\quad ( \forall u \in H)
\label{eq15}
\end{align}
Also, it is clear that
\begin{align}
&
\langle u \otimes s, [A_1 \otimes I, A_2 \otimes I](u \otimes s) \rangle 
\nonumber
\\
=
&
\langle u , [A_1 , A_2 ]u  \rangle
\quad ( \forall u \in H)
\label{eq16}
\end{align}

Summing up ((\ref{eq10}),(\ref{eq12}),(\ref{eq14}),(\ref{eq15}),(\ref{eq16})), we can conclude that
\begin{align}
&
{\Delta}_{\widehat{N}_1}^{u \otimes s} \cdot { \Delta}_{\widehat{N}_2}^{u \otimes s}
(=
{\overline \Delta}_{\widehat{N}_1}^{u \otimes s} \cdot {\overline \Delta}_{\widehat{N}_2}^{u \otimes s}
)
\nonumber
\\
\ge
&
\frac{1}{2}
| \langle u ,
[A_1,A_2]  u  \rangle |
\quad ( \forall u \in H \mbox{ such that } ||u||=1 )
\label{eq17}
\end{align}
which is Ishikawa's formulation of Heisenberg's uncertainty principle ({\rm cf.} \cite{Ishi1}).

\par
\noindent
{\bf Remark 3} Assume that $[A_1, A_2]= \hbar {\sqrt{-1}}I$. If Hypothesis 1 is not assumed, we can say in what follows. That is, for any positive $\epsilon$, there exist $s \in K (\|s \|=1)$, ${\widehat{A}}_i (i=1,2)$, $u \in H (\| u \|=1)$ such that
\begin{align*}
\Delta_{\widehat{N}_1}^{u \otimes s} < \epsilon, \quad \Delta_{\widehat{N}_2}^{u \otimes s} < \epsilon
\end{align*}
(cf. Remark 3 in  ref.\cite{Ishi1}). Thus, if we hope that Heisenberg uncertainty principle (\ref{eq17}) holds, the same average condition is indispensable.

\par
\noindent
\subsection{\normalsize
Heisenberg uncertainty principle without the same average condition
}
We believe that Hypothesis 1 is very natural. However, 
in this section, we do not assume Hypothesis 1 ( the same average condition).

Put $\sigma (A_i;u)=\| (A - \langle u, A_iu \rangle )u \|$. Using the Robertson uncertainty relation, we can estimate the second term (or, precisely,
$\langle u \otimes s,$"the second term"$(u \otimes s) \rangle$) 
in (\ref{eq9}) as follows:
\begin{align}
&
2 {\overline \Delta}_{\widehat{N}_1}^{u \otimes s} \cdot \sigma(A_2;u)
\ge
|\langle u \otimes s, [{\widehat N}_1, A_2 \otimes I](u \otimes s) \rangle| 
\nonumber
\\
&
\qquad  ( \forall u \in H \mbox{ such that } ||u||=1)
\label{eq18}
\end{align}
Similarly, we estimate the third term in (\ref{eq9}) as follows:
\begin{align}
&
2 {\overline \Delta}_{\widehat{N}_2}^{u \otimes s} \cdot  \sigma(A_1;u)
\ge
|\langle u \otimes s, [A_ \otimes I, {\widehat N}_2](u \otimes s) \rangle |
\nonumber
\\
&
\qquad  ( \forall u \in H \mbox{ such that } ||u||=1)
\label{eq19}
\end{align}
Summing up ((\ref{eq8}),(\ref{eq10}), (\ref{eq16}),(\ref{eq18}),(\ref{eq19})), we can conclude that
\begin{align}
& { \Delta}_{\widehat{N}_1}^{u \otimes s} \cdot { \Delta}_{\widehat{N}_2}^{u \otimes s}
+{ \Delta}_{\widehat{N}_2}^{u \otimes s} \cdot \sigma(A_1;u)
+{ \Delta}_{\widehat{N}_1}^{u \otimes s} \cdot \sigma(A_2;u)
\nonumber
\\
\ge
&
{\overline \Delta}_{\widehat{N}_1}^{u \otimes s} \cdot {\overline \Delta}_{\widehat{N}_2}^{u \otimes s}
+{\overline \Delta}_{\widehat{N}_2}^{u \otimes s} \cdot \sigma(A_1;u)
+{\overline \Delta}_{\widehat{N}_1}^{u \otimes s} \cdot \sigma(A_2;u)
\nonumber
\\
\ge
&
\frac{1}{2}
| \langle u ,
[A_1,A_2]  u  \rangle |
\quad ( \forall u \in H \mbox{ such that } ||u||=1)
\label{eq20}
\end{align}
Since Hypothesis 1 is not assumed in this section, it is a matter of course that this (\ref{eq20}) is more rough than the (\ref{eq17}).

\par
\noindent
{\bf Remark 4} (Ozawa's inequality).  In \cite{Oza, Oza2}, M. Ozawa tried to formulate Heisenberg's $\gamma$-ray microscope thought experiment in his interpretation, and proposed the following inequality (so called Ozawa's inequality):
\begin{align}
&
\epsilon (A_1) \eta(A_2) + \eta(A_2) \sigma (A_1)  +\epsilon (A_1) \sigma(A_2)
\nonumber
\\
\ge
&
\frac{1}{2}
| \langle u ,
[A_1,A_2]  u  \rangle |
\label{eq21}
\end{align}
which is, by our notation, rewritten as follows.
\begin{align}
& { \Delta}_{\widehat{N}_1}^{u \otimes s} \cdot { \Delta}_{\widehat{N}_2}^{u \otimes s}
+{ \Delta}_{\widehat{N}_2}^{u \otimes s} \cdot \sigma(A_1;u)
+{ \Delta}_{\widehat{N}_1}^{u \otimes s} \cdot \sigma(A_2;u)
\nonumber
\\
\ge
&
\frac{1}{2}
| \langle u ,
[A_1,A_2]  u  \rangle |
\quad ( \forall u \in H \mbox{ such that } ||u||=1)
\label{eq22}
\end{align}
Note that this (\ref{eq22}) is mathematically the same as the above (\ref{eq20}), but the (\ref{eq21}) is not. Here, it should be noted that Ozawa's assertion is just the (\ref{eq21}), that is,
\begin{itemize}
\item[(K)]the physical meanings of "error"($\epsilon (A_1)(= { \Delta}_{\widehat{N}_1}^{u \otimes s})$) and "disturbance"$\eta(A_2)(= { \Delta}_{\widehat{N}_1}^{u \otimes s})$ are distinguished in Ozawa's inequality (\ref{eq21}).
\end{itemize}
Therefore there is a great gap between Ozawa's inequality (\ref{eq21}) and the (\ref{eq20}). In fact, the (\ref{eq20}) is not the mathematical representation of Heisenberg's $\gamma$-ray microscope thought experiment.
Now we think that it may be impossible to formulate this (K) in the linguistic interpretation, since the (E$_2$) says that anything after measurement can not be described. In fact, we are not successful yet. Therefore, we consider that the other interpretation is indispensable for the understanding of Ozawa's inequality (\ref{eq21}), and thus, the (\ref{eq20}) and the (\ref{eq21}) are different assertions.

\par
\noindent
\par
\noindent
\section{\large
Quantum Zeno effects
}
\par
\noindent
Let
$[B_c(H),B(H)]_{B(H)}$ 
be the basic structure.
Let ${\mathbb P}=[P_n ]_{n=1}^\infty$ be the spectral resolution in $B(H)$, that is,@for each $n$, $P_n \in B(H)$ is a projection such that
\begin{align*}
\sum_{n=1}^\infty P_n =I
\end{align*}
Define the $(\Psi_{\mathbb P})_*: Tr(H) \to Tr(H)$ 
such that
\begin{align*}
(\Psi_{\mathbb P})_* (|u \rangle \langle u |) = \sum_{n=1}^\infty |P_n u \rangle \langle P_n u |
\quad (\forall u \in H)
\end{align*}
Also, we define the Schr\"{o}dinger time evolution $(\Psi_S^{\Delta t})_* : Tr(H) \to Tr(H)$  such that
\begin{align*}
(\Psi_S^{\Delta t})_* (|u \rangle \langle u |) = |e^{-\frac{i {\cal H} \Delta t}{\hbar}}u \rangle \langle e^{-\frac{i {\cal H} \Delta t}{\hbar}} u |
\quad (\forall u \in H)
\end{align*}
Consider $t=0,1$. Putting $\Delta t = \frac{1}{N}$, $H=H_0=H_1$, we can define the
$(\Phi_{0,1}^{(N)})_*: Tr(H_0) \to Tr(H_1)$ 
such that
\begin{align*}
(\Phi_{0,1}^{(N)})_* =((\Psi_S^{1/N})_* (\Psi_{\mathbb P})_*)^N
\end{align*}
which induces the Markov operator
$\Phi_{0,1}^{(N)} : B(H_1) \to B(H_0)$ as the dual operator
$\Phi_{0,1}^{(N)} =((\Phi_{0,1}^{(N)})_*)^*$.
Let $\rho=|\psi \rangle \langle \psi |$ be a state at time $0$.
Let $ {\mathsf{O}_1}{\; :=} (X, {\cal F}, F)$ be an observable in $B(H_1)$.
Thus, we have a measurement: 
$$
{\mathsf{M}}_{B(H_0)} (\Phi_{0,1}^{(N)} {\mathsf{O}_1}, S_{[\rho]})
$$
$\big($
or more precisely,
${\mathsf{M}}_{B(H_0)} (\Phi_{0,1}^{(N)}{\mathsf{O}}{\; :=} (X, {\cal F}, \Phi_{0,1}^{(N)}F),$
$ S_{[|\psi \rangle \langle \psi |]})$
$\big)$.
Here, Axiom 1 says that
\begin{itemize}
\item[(L)] the probability that the measured value obtained by the measurement belongs to
$\Xi (\in {\cal F})$ is given by
\begin{align}
tr(| \psi \rangle \langle \psi | \cdot \Phi_{0,1}^{(N)}F(\Xi))
\label{eq23}
\end{align}
\end{itemize}

Now we shall explain "quantum Zeno effect" in the following example.
\par
\noindent
{\bf Example 1} Let $\psi \in H$ such that $\|\psi \|=1$.
Define the spectral resolution 
\begin{align}
{\mathbb P}=[ P_1 (=|\psi \rangle \langle \psi |), P_2(=I-P_1) ]
\label{eq24}
\end{align}
And define the observable $ {\mathsf{O}_1}{\; :=} (X, {\cal F}, F)$ in $B(H_1)$ such that
$$
X=\{ x_1 , x_2 \}, \qquad {\cal F}=2^X
$$
and
$$
F(\{x_1 \})=|\psi \rangle \langle \psi |(=P_1),
F(\{x_2 \})=I- |\psi \rangle \langle \psi |(=P_2),
$$
Now we can calculate (\ref{eq23})(i.e., the probability that a measured value $x_1$ is obtained) as follows.
\begin{align}
(\ref{eq23}) &=
\langle \psi,
((\Psi_S^{1/N})_* (\Psi_{\mathbb P})_*)^N (|\psi \rangle \langle \psi |)
\psi
\rangle
\nonumber
\\
&
\ge
|\langle \psi , e^{-\frac{i {\cal H} }{\hbar N}}\psi \rangle
\langle \psi , e^{\frac{i {\cal H} }{\hbar N}}\psi \rangle|^N
\nonumber
\\
&
\approx
\big(1 - \frac{1}{N^2} ( || (\frac{ {\cal H} }{\hbar }) \psi ||^2 - |\langle \psi,  (\frac{ {\cal H} }{\hbar }) \psi \rangle |^2) \big)^N \to 1 
\nonumber
\\
&
\qquad \qquad \qquad \qquad ( N \to \infty)
\label{eq25}
\end{align}
Thus, if $N$ is sufficiently large, we see that
\begin{align*}
{\mathsf{M}}_{B(H_0)} (\Phi_{0,1}^{(N)} {\mathsf{O}_1}, S_{[|\psi \rangle \langle \psi |]})
=
{\mathsf{M}}_{B(H_0)} (\Phi_I  {\mathsf{O}_1}, S_{[|\psi \rangle \langle \psi |]})
\\
(\text{where $\Phi_I:B(H_1) \to B(H_0)$ is the identity map})
\end{align*}
or, we say, roughly speaking in terms of the Schr\"{o}dinger picture, that the state $|\psi \rangle \langle \psi |$ does not move. 

\par
\noindent
{\bf Remark 5} The above argument is motivated by B. Misra and E.C.G. Sudarshan \cite{Misr}. However, the title of their paper: "The Zeno's paradox in quantum theory" urges us to guess that
\begin{itemize}
\item[(M)] the spectral resolution ${\mathbb P}$ of (\ref{eq24}) is regarded as an observable (or moreover, measurement) in their paper \cite{Misr}.
\end{itemize}
If this (M) is their assertion, we can not understand "quantum Zeno effect". That is because the linguistic interpretation require the commutative condition (4) should be satisfied, however, ${\mathbb P}$ and $\Psi_S^{\Delta t}{\mathbb P}$ do not commute.
In the sense of Example 1, this effect should be called "brake effect" and not "watched pot effect".


\par
\noindent
\section{\large
Conclusions
}
\par
\noindent

In this paper, we point out the possibility that
\begin{itemize}
\item[(N)] two nice ideas (K) and (M) can not be understood in the linguistic interpretation.
\end{itemize}
That is because our theory is not concerned with any influence after a measurement.
In spite of the difficulty such as (N), we do not give up to assert the linguistic interpretation, since it has a great power of description ({\it cf.} refs.
{{{}}}{\cite{Ishi2}-\cite{Ishi11}}). 

It is always interesting to find a phenomenon that can not be explained in MT.
Thus, in spite of our conjecture (N), we earnestly hope that the readers investigate the following problem:  
\begin{itemize}
\item[(O)] Describe Ozawa's inequality (\ref{eq21}) in the linguistic interpretation!         
\end{itemize}                                                                   This problem is very important in quantum mechanics. That is because it is generally believed that the difference of interpretations is usually negligible in practical problems.  If the formulation of Heisenberg uncertainty principle depends on quantum interpretations, our next problem may be to investigate "What is the most certain interpretation?"  And we believe that the linguistic interpretation is quite hopeful. However, it should be examined from various points of view.

\rm
\par
\renewcommand{\refname}{
\large 
References}
{
\small

\normalsize
}

\end{document}